\documentclass{sf2a-conf}
\usepackage{graphicx}
\begin{document}

\TitreGlobal{SF2A 2006}

\title{Is Dark Energy Abnormally Weighting?}

\author{F\"uzfa, A.}\address{F.N.R.S. post-doctorate researcher, GAMASCO, University of Namur (F.U.N.D.P.), Belgium}
\author{Alimi, J.-M.}\address{Laboratory Universe and Theories, CNRS UMR 8102,\\
 Observatoire de Paris-Meudon and Universit\'e Paris VII, France}
\runningtitle{Is Dark Energy Abnormally Weighting?}
\setcounter{page}{237}
\index{F\"uzfa, A.}
\index{Alimi, J.-M.}

\maketitle
\begin{abstract} 
We present a new interpretation of dark energy in terms of an \textit{Abnormally Weighting Energy} (AWE). This means
that dark energy does not couple to gravitation in the same way as ordinary matter, yielding a violation of the weak and 
strong equivalence principles on cosmological scales. The resulting cosmological mechanism accounts for the Hubble diagram 
of type Ia supernovae in terms of both cosmic acceleration and variation of the gravitational constant while still accounting
for the present tests of general relativity. This explanation 
allows to build dark energy models (i) without violation of the strong energy condition $p<-\rho c^2/3$ 
(ii) with non-negligible direct couplings to gravitation and (iii) natural convergence mechanism toward general relativity.
\end{abstract}
%
\section{Motivations for an abnormally weighting energy}
Maybe the most helpful astrophysical evidence for determining the nature of dark energy is the dimmed magnitude of type Ia supernovae
(see Astier et al. 2006 and references therein).
This tell us that either the recent cosmic expansion has accelerated or (and) these objects are not standard candles, which could
have happened if, for instance, fundamental constants of physics have recently varied.
This is the first explanation that is retained in the so-called concordance model $\Lambda CDM$ 
in which a positive cosmological constant constitutes
the dark energy sector of cosmology. This convenient explanation does not require to give up neither general relativity (GR) nor
the cosmological principle and furthermore allows to account for dark energy effects with only one additional parameter. 
However, the cosmological constant is usually interpreted as the stress-energy term associated to the vacuum 
state which has non-vanishing energy due to quantum fluctuations. 
Doing so, one recovers the old problem of the overestimation of the cosmological constant. Indeed,
if one assumes that the cut-off energy scale at which GR is no longer valid is the Planck length, then the vacuum energy density 
acting as a cosmological constant is $\rho_{\Lambda}\approx m_{Pl}^4\approx 10^{76} GeV^4$
(in geometrical units where $\hbar=c=1$, $G=m_{Pl}^{-2}$). Which mechanism can be so precise to make such a huge quantity vanishing?
This constitutes the so-called \textit{fine-tuning problem} of the cosmological constant. What is worst now with the astrophysical
evidences for the cosmological constant, is that the vacuum energy density is indeed not vanishing but is of cosmological order of magnitude:
$\rho_{\Lambda}\approx \rho_{c}(z=0)\approx 10^{-47} GeV^4\cdot$.
We have therefore a huge disagreement between theory and experiment of more than $120$ orders of magnitude! 
Why is the vacuum energy density, 
something that should be set once for all in the universe and should depend on quantum mechanics, is of the order of 
the critical density today? Such a value suggests a cosmological mechanism behind the value of the cosmological constant...
Even worst, this means that we live in a very particular epoch in the history of the universe : the time when the cosmological 
constant, whose value has \textit{a priori} nothing to do with cosmology if it is fixed by quantum considerations, starts dominating the 
energy content of the universe. This constitutes the so-called \textit{coincidence problem} associated to the cosmological constant.
However, one should keep in mind that this interpretation of dark energy in terms of \textit{weighting} quantum fluctuations
is obviously a quantum gravity problem. One can therefore expect that the final solution to this problem 
might well be quite different from the simple and maybe rather naive cosmological constant.
An interesting way-out of these problems (Padmanabhan 2005) is to consider that dark energy is not related to the absolute 
value of the energy of the quantum ground state but rather to fluctuations around it. For instance, one can links the dark energy density
to the size of the particle (Hubble) horizon which could solve the coincidence problem. However, the resulting mechanism can no longer
be interpreted as a cosmological constant as the horizon is not an absolute, constant, notion...\\
\\
Another widespread interpretation of dark energy is that of quintessence, a "\textit{varying cosmological constant}" which provide
a cosmological mechanism aimed to solve coincidence (and in some cases fine-tuning) problems. This most of time consists on considering
dark energy as a self-interacting scalar field which couples mostly to ordinary matter indirectly, i.e. through purely gravitational couplings.
This means that quintessence does only modify cosmic expansion and does not alter conventional physics.
While rolling down its self-interacting potential, this scalar field might produce sufficiently negative pressures ($p<-\rho c^2/3$)
to provide the necessary cosmic acceleration. In the cases of tracking potentials, there exists attractors in the scalar field dynamics which
ensure a solution to both fine-tuning and coincidence problems.\\
\\
Let us now consider what happens if dark energy couples directly to ordinary matter, which is obviously not possible 
for the cosmological constant. According to its "\textit{darkness}", quintessence is a neutral field and the non-gravitational
interactions it could have with matter will only affects the coupling constants. Such a non-minimally coupled scalar field, sometimes
called "\textit{extended quintessence}" therefore yields a violation of the equivalence principle. In the case
where the scalar field affects the couplings to the metric and therefore
makes the gravitational constant varying, we face the famous \textit{scalar-tensor} theories of gravitation.
However, if we consider that the $70\%$ of missing energy is the whole contribution of such a non-minimally coupled
scalar field, then it could be difficult to match the present tests
of GR without assuming its direct couplings to be \textit{extremely} weak. Even worst, as soon as these couplings are non-vanishing, the endless domination of a non-minimally coupled scalar
field acting as dark energy might inescapably lead to a disastrous violation of the equivalence principle in the future.
Indeed, an unbounded scalar field might drive the coupling constants to values arbitrary far from GR. 
Therefore, if dark energy is not completely dark and if it does not rule temporarily the cosmic expansion, then GR itself
could be a coincidence of the present universe... In this way, the non-gravitational couplings of dark energy to matter are extremely important and should
be considered carefully. \\
\\
Another possibility of such couplings is that 
"dark" energy violates the weak
equivalence principle (WEP) on cosmological scales. This means that it does not couple to
gravitation as usual matter does and weights abnormally. Doing so, its related gravitational
energy will be felt differently by other types of matter, therefore
violating also the strong equivalence principle (SEP). Under this
assumption, we have built a plausible dark energy mechanism (i) without violation of the strong energy condition (SEC) $p<-\rho c^2/3$ 
(ii) with non-negligible direct couplings to gravitation and (iii) with a 
natural convergence mechanism toward general relativity (F\"uzfa \& Alimi 2006a hereafter AWEI ; F\"uzfa \& Alimi 2006b hereafter AWEII).
\section{An AWE Born-Infeld Gauge Interaction}
The AWE hypothesis consists on dividing the energy
content of the universe into three parts : a
gravitational sector described by pure spin 2 (graviton) and spin 0
(dilaton) degrees of freedom, a matter sector containing the usual
fluids of cosmology (baryons, photons, dark matter, ...)
and an AWE sector. The violation of the WEP by the AWE 
can be represented by different couplings between gravity,
the AWE and usual matter:
\begin{eqnarray}
\label{action1}
S&=&\frac{1}{2\kappa}\int\sqrt{-g}d^4x\left\{R-2g^{\mu\nu}\partial_\mu\varphi\partial_\nu\varphi\right\}
+S_{awe}\left[\psi_{awe},A_{awe}^2(\varphi)g_{\mu\nu}\right]
+S_m\left[\psi_m,A_m^2(\varphi)g_{\mu\nu}\right],
\end{eqnarray}
where $\kappa$ is the "\textit{bare}" gravitational coupling
constant. Similar effective theories of gravitation with
non-universal couplings are also motivated by string theory (see Damour \& Polyakov 1994
and references therein). In the previous action, $g_{\mu\nu}$ is the Einstein metric,
$\varphi$ is a gravitational scalar field, $S_{awe}$ is the action for the AWE
sector with fields $\psi_{awe}$ and $S_m$ is the usual matter
sector with matter fields $\psi_m$; $A_{awe}(\varphi)$ and $A_m(\varphi)$ being the coupling functions 
to the metric $g_{\mu\nu}$ for the AWE and matter sectors respectively. The non-universality of the
gravitational couplings ($A_{awe}\ne A_m$) yields a violation of the
WEP: experiments using the new
AWE sector would provide a different inertial mass
than all other experiments. The action (\ref{action1}) is written
in the so-called "\textit{Einstein frame}" of the physical degrees of freedom. In this frame, 
the metric
components are measured by using purely gravitational rods and
clocks, i.e. not build upon any of the matter fields nor the ones from the AWE sector. \\
\\
The frame in which cosmology has to be done, the "\textit{Dicke-Jordan}" observable frame,
is given
by the conformal transformation
\begin{equation}
\label{obsframe} \tilde{g}_{\mu\nu}=A^2_m(\varphi)g_{\mu\nu}
\end{equation}
using the coupling function to ordinary matter. Indeed, observations are made
thanks to ordinary matter and, in this
observable frame, the metric $\tilde{g}_{\mu\nu}$ couples universally to
ordinary matter and is measured by clocks and rods made of usual
matter (but not build upon the AWE sector). The violation
of the WEP therefore only concerns the new
AWE sector that was introduced in (\ref{action1}). Throughout this
paper, quantities with a tilde will refer to the observable frame
given
by (\ref{obsframe}).\\
\\
Let us focus here on the dynamics of the scalar gravitational field, which is ruled by
the following equation :
\begin{eqnarray}
\label{kg}
\ddot{\varphi}+3\frac{\dot{a}}{a}\dot{\varphi}+\frac{\kappa}{2}\alpha_{awe}(\varphi)\left(\rho_{awe}-3p_{awe}\right)
+\frac{\kappa}{2}\alpha_m(\varphi)\rho_m=0,
\end{eqnarray}
where $\alpha_i(\varphi)=d\ln A_i(\varphi)/d\varphi\cdot$
The AWE sector appears as an additional driving term in the scalar dynamics that competes with 
ordinary matter (dust in (\ref{kg})). A possible candidate for dark energy can therefore be
an AWE type of matter which has recently altered the well-known convergence mechanism toward general relativity during the
matter-dominated epoch (Damour \& Nordtvedt 1993 ; Serna, Alimi \& Navarro 2002) due to its violation of the WEP. Such a recent modification
of GR can result in
an accelerated expansion and a variation of the gravitational constant.
\\
\\
We can now particularize the general mechanism by specifying the AWE sector through its equation of state (EOS). In 
AWEI and AWEII, we have chosen a gauge interaction of Born-Infeld type (see also Dyadichev et al. 2002) as a source
of WEP violation. Such a gauge interaction is characterized by an EOS $\omega_{awe}=p/\rho$ varying between 
$-1/3$ (Nambu-Goto string gas) at high-energies and $1/3$ (radiation) in the low-energy regime corresponding to Yang-Mills 
dynamics. This yields to four important properties for this particular AWE mechanism. First, the SEC is never violated in the Einstein frame, and
the cosmic acceleration is produced while moving to the observable frame. Second,
the negative EOS at high-energies ensures a late occurence of dark energy, without modification 
of the physics in the radiative era. Third, this particular EOS results in a natural transient dark energy mechanism
and an ensured convergence toward GR. Last but not least, the final radiative EOS allows to have a dominant AWE today 
($\Omega_m<\Omega_{awe}$) in the same time as a gravitational theory close to GR. Indeed, once in the radiative regime $\omega_{awe}=1/3$, 
the AWE Born-Infeld interaction decouples from the scalar field (see (\ref{kg})) and let the ordinary matter accomplishing his work to re-establish GR
within the stringent constraints we know about the equivalence principle,
even though the AWE still dominates temporarily the energy content of the Universe. \\
\\
Let us examine closely some key features of the AWE mechanism 
and refer the reader to our paper AWEII for more details.
On Fig. 1 (left),
the reader will find the Hubble diagram of type Ia supernovae (data from Astier et al. 2006)
together with the theoretical predictions from the concordance model and the model presented
here. For the last, the distance moduli relation contains an additional term from a toy model allowing to
account for the variation of the Newton constant on supernovae physics (Gaztanaga et al., 2001;
Riazuelo \& Uzan, 2002). Under these assumptions, we found that the AWE and concordance models both accounts fairly for data
although the first contains more free parameters than the concordance model. Furthermore, the model presented in Fig. 1
also accounts for the stringent constraints on the PPN parameters $\gamma$ and $\beta$ as well as on the absolute time
variation of $G$ (see AWEII). The central part of Fig. 1 illustrates the evolution of the effective gravitational constant and the 
observable acceleration factor
$\tilde{q}=1+\left(d\tilde{H}/d\tilde{t}\right)/\left(\tilde{H}^2\right)$ that lead to the prediction of Fig. 1 (left).
The action of the Born-Infeld AWE provides both cosmic acceleration and a variation of $G$ of about 70\% at $z=1$ to match data.
Finally, one will find in Fig. 1 (right) the evolution of the energy densities for the protagonists of the AWE mechanism (AWE,
matter and the gravitational scalar field). 
First, the AWE gauge field starts sub-dominant at
the end of the radiation-dominated era while gravitation is
well described by GR. Then, as the energy densities progressively cool down to coincidence
the scalar field is pushed away from GR by the increasing
repulsive influence of the AWE. This repulsive
influence rapidly decreases
as the gauge field becomes radiative and
decouples from the scalar sector. Between this period and today,
matter becomes the dominant driving term and
attracts towards GR to finally achieve the level of precision we
know for it today. This occurs when the AWE sector is still dominating the present energy content
of the Universe as previously explained. During a short period of time
in the very recent cosmic history, gravitation was substantially
different from GR and led to dark energy effects we observe in the Hubble diagram of type Ia supernovae. 
\section{Conclusion}
The original AWE hypothesis we presented here assumes that dark energy violates the WEP on cosmological scales,
which alters the well-known convergence mechanism of scalar-tensor theories toward GR during the matter-dominated era.
We have shown in AWEII that this assumption allows to account for the Hubble diagram of type Ia supernovae through
cosmic acceleration and time variation of the gravitational constant, account for stringent constraints on the equivalence principle
today, and does not require too negative pressures ($p\ge -\rho/3$).
Furthermore, this assumption is well-motivated as a generalization of scalar-tensor theories for which
the WEP is usually assumed. AWE can also represent an interesting alternative to usual
dark energy models in which this extra gravitational energy content is extremely weakly coupled to
matter therefore producing very low variation of fundamental constants and violates the SEC. 
At least, the AWE assumption can help us to test these key features of
dark energy.
As well, this hypothesis is also expected to be well constrained by post-recombination on the cosmic
microwave background and large-scale structure formation due to the modified recent cosmic expansion and the 
rather important variation of $G$ it could predict. It should also
modify the universality of free fall, especially on cosmological scales where the amount of AWE
is not negligible.
All these effects therefore constitute very important clues for our quest to determine whether some processes in
the universe are not ruled by the equivalence principle and how far Einstein was right in postulating it. 
For all these reasons, we claim that the AWE hypothesis should deserve a particular attention in future works 
as it goes well beyond the scope of finding a valuable explanation of dark energy 
but directly questions the deep nature of gravitation itself.
\begin{figure}
\begin{center}
\begin{tabular}{ccc}
\includegraphics[scale=0.3]{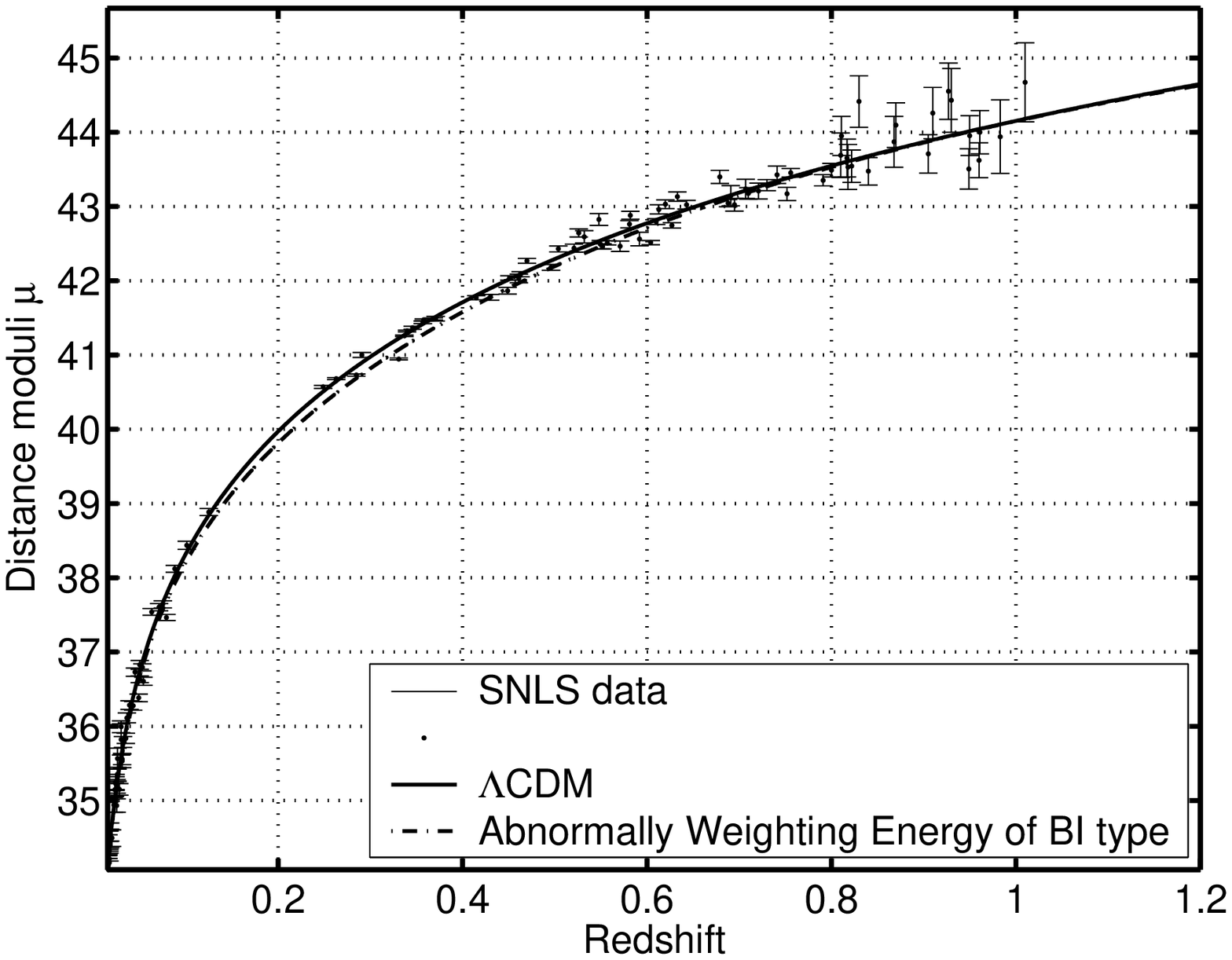} &
\includegraphics[scale=0.3]{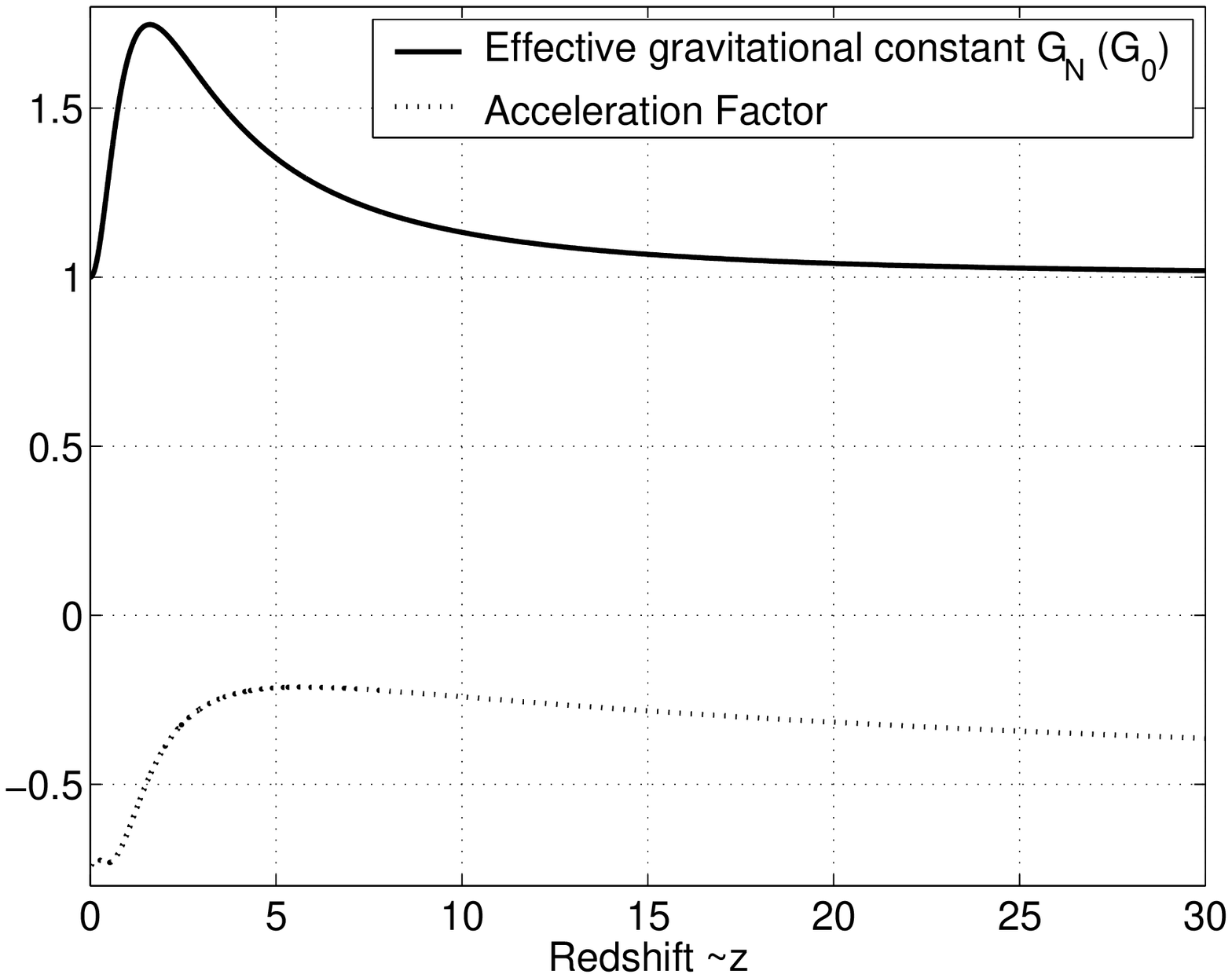}
\includegraphics[scale=0.3]{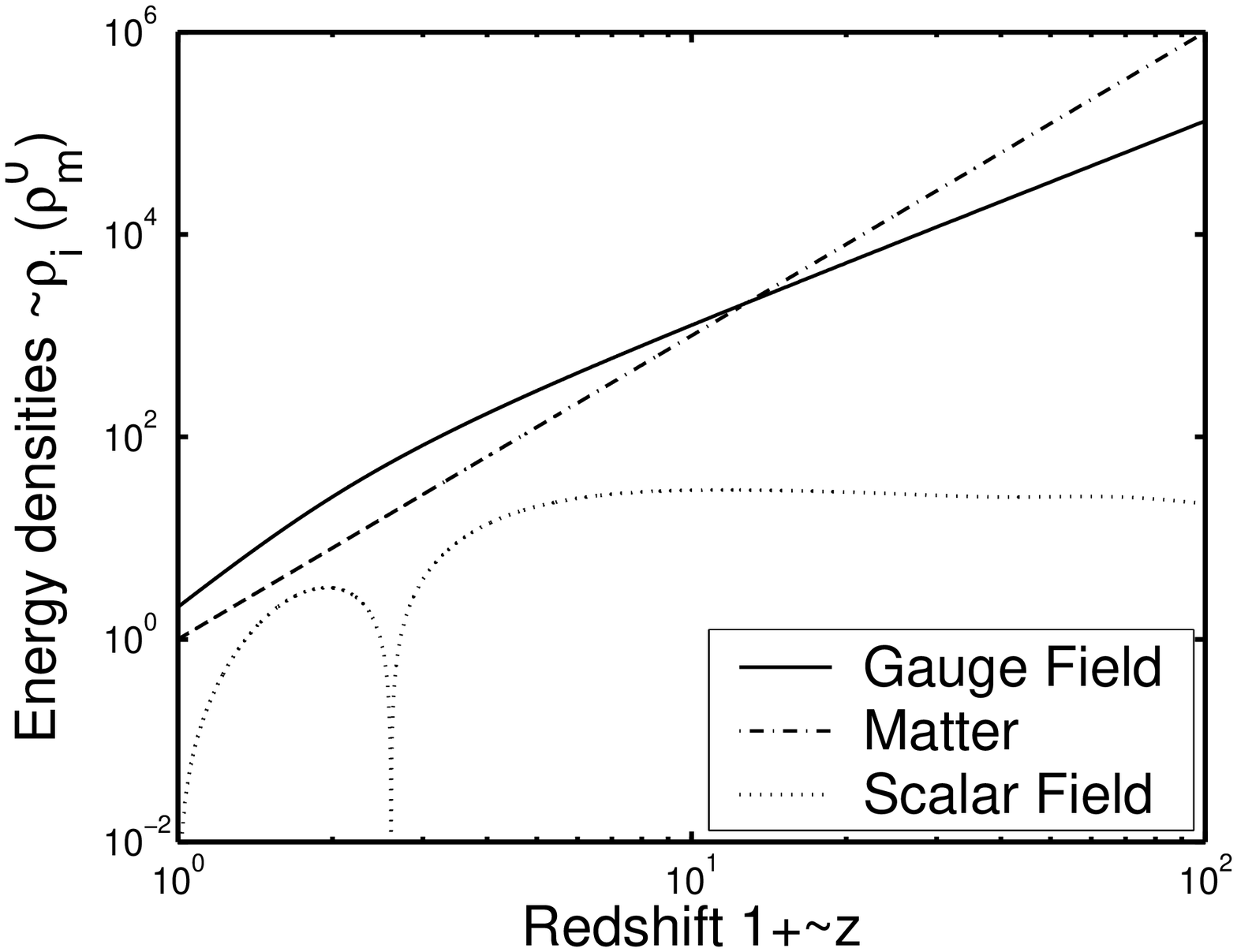}
\end{tabular}
\end{center}
\caption{(left) Hubble diagram of SNLS 1st year data set with the best fit
$\Lambda CDM$ flat model (solid line, $\Omega_m(a_0)=0.26$,
$\bar{\chi}^2/dof=1.03$) and the AWE model (dash-dotted line,
$\bar{\chi}^2/dof=1.09$) ($H_0=70 km/s/Mpc$); (center) cosmological evolutions of $G_N$ (in units of
bare $G$), the cosmic acceleration factor and (right) the energy densities with the redshift} \label{fig1}
\end{figure}



\end{document}